\documentclass[a4paper,12pt]{article}
\pdfoutput=1 
\usepackage{jheppub} 
\usepackage[T1]{fontenc} 
\usepackage{amsfonts,amssymb,amsmath}
\usepackage{color}

\newcommand{\be}{\begin{equation}}
\newcommand{\ee}{\end{equation}}
\newcommand{\bea}{\begin{eqnarray}}
\newcommand{\eea}{\end{eqnarray}}

\title{The fate of the Higgs vacuum }

\author[a,b]{Philipp Burda\footnote{On leave of absence from ITEP, Moscow.}}
\author[b,c]{Ruth Gregory}
\author[d]{Ian G.\ Moss}

\affiliation[a]{Racah Institute of Physics, Hebrew University, Jerusalem 91904, Israel}
\affiliation[b]{Centre for Particle Theory, Durham University,
South Road, Durham, DH1 3LE, UK}
\affiliation[c]{Perimeter Institute, 31 Caroline Street North, Waterloo, 
ON, N2L 2Y5, Canada}
\affiliation[d]{School of Mathematics and Statistics, Newcastle University, 
Newcastle Upon Tyne, NE1 7RU, U.K.}

\emailAdd{philipp.burda@mail.huji.ac.il}
\emailAdd{r.a.w.gregory@durham.ac.uk}
\emailAdd{ian.moss@newcastle.ac.uk}

\abstract{
We have recently suggested that tiny black holes can act as nucleation
seeds for the decay of the metastable Higgs vacuum. Previous results 
applied only to the nucleation of thin-wall bubbles, and covered a very 
small region of parameter space. This paper considers bubbles of arbitrary
profile and reaches the same conclusion: black holes seed rapid vacuum decay. 
Seeded and unseeded nucleation rates are compared, and 
the gravitational back reaction of the bubbles is taken into account.
The evolution of the bubble interior is described for the unseeded nucleation.
Results are presented for the renormalisation group improved Standard 
Model Higgs potential,
and a simple effective model representing new physics.
}

\preprint{DCPT-16/01}

\begin{document}

\maketitle

\section{Introduction}

Although many phase transitions in physical models are second order, our intuitive 
picture of a phase transition is determined by our most common experience: boiling 
water. Such a first order phase transition proceeds by nucleation of bubbles of the new 
phase, often around impurities, which then expand. This intuitive picture of a first 
order phase transition has a corresponding physical and mathematical analogy
in quantum phase transitions between different vacua
\cite{coleman1977,callan1977,CDL,Kobzarev:1974cp}. Such decay processes have
current relevance due to the possible metastability of the Higgs vacuum
\cite{Degrassi:2012ry,Buttazzo:2013uya,Gorsky:2014una,Bezrukov:2014ina,Ellis:2015dha,Blum:2015rpa},
mooted some time ago \cite{Krive:1976sg,Cabibbo:1979ay,1982Natur.298,
Lindner:1988ww,Sher:1988mj,Isidori:2001bm,Espinosa:2007qp,
Isidori:2007vm,EliasMiro:2011aa}, but lent recent credence by the measured value
of the Higgs mass \cite{ATLAS:2012ae,Chatrchyan:2012tx}.

The nucleation of a bubble of a different vacuum phase was described
in a series of papers by Coleman and collaborators 
\cite{coleman1977,callan1977, CDL}, in which a Euclidean approach is
used to describe the leading order contribution to the wavefunction for
decay. For vacua separated by large barriers, this is well approximated
by assuming the two vacua are separated by relatively thin wall of energy,
throughout which the fields vary from one vacuum to the other. The 
gravitational effect of this ``thin-wall'', as well as of the corresponding
vacuum energies, can be computed precisely in (Euclidean) Einstein
gravity \cite{GibHawk}, using the Israel equations \cite{Israel} to model the bubble wall. 
Coleman and de Luccia \cite{CDL} described this physical picture
of vacuum decay in the universe, and presented the Coleman-de Luccia
(CDL) instanton, which is now the ``gold standard'' for describing vacuum 
decay.

The single instanton picture of Coleman et al.\ is however extremely idealised. 
There are no features to the solution other than the bubble -- in particular, 
no description of inhomogeneities. Given the gravitational set-up of CDL,
the most natural and simplest inhomogeneity to introduce is a black hole,
and although early work did explore this \cite{Hiscock,Berezin:1987ea,Berezin:1990qs}, it
failed to properly account for the impact of the conical deficits that 
inevitably arise in the Euclidean calculations.  In \cite{GMW}, the 
effect of said conical deficits was carefully computed, and a potentially
large enhancement of the CDL rates was demonstrated in the context
of tunnelling from a positive to zero cosmological constant. 
(See also \cite{Aguirre:2005xs} for a study of general thin wall
solutions.)

Applying these ideas to the Higgs vacuum, in \cite{BGM1,BGM2} 
we recently provided a proof of principle that the lifetime of the vacuum could 
become precipitously short in the presence of primordial black
holes, paralleling the intuition of impurities catalysing a phase transition.
However, the semi-analytic arguments
we used (based on the Israel ``thin-wall'' formalism \cite{Israel}) meant
that we could only apply these conclusions to a very small and artificial
region of parameter space within a (quantum gravity) corrected Higgs 
potential. 
In \cite{BGM1}, we provided preliminary evidence that this parameter 
space restriction was an artefact of the constraints imposed on the 
potential by demanding that it allow a thin-wall approximation for the
instanton. The purpose of this paper is to confirm and flesh out
this claim: Specifically, by integrating out the coupled Einstein-Higgs 
equations of motion for a Euclidean instanton solution, we will show that
for a wide range of BSM / quantum gravity corrections (or indeed none at all!)
to the Higgs potential, the presence of a micro-black hole can prove lethal to our
universe.

\section{``Standard'' Higgs vacuum decay}

Before embarking our presentation, we first briefly review the standard
description of vacuum decay. We discuss the simplified parametrisation of
the Higgs potential we will be using in our integrations, then discuss briefly
the usual CDL-type instanton, however, rather than approximate this by
the Israel-thin-wall description (followed by CDL), we compute this
instanton numerically. This generalises
previous results on the instanton solutions in flat space 
\cite{Branchina:2013jra,Branchina:2014rva} and semi-analytical
results in de Sitter space \cite{Shkerin:2015exa}.

\subsection{The Higgs potential}

\begin{figure}[htb]
\begin{center}
\includegraphics[width=0.8\textwidth]{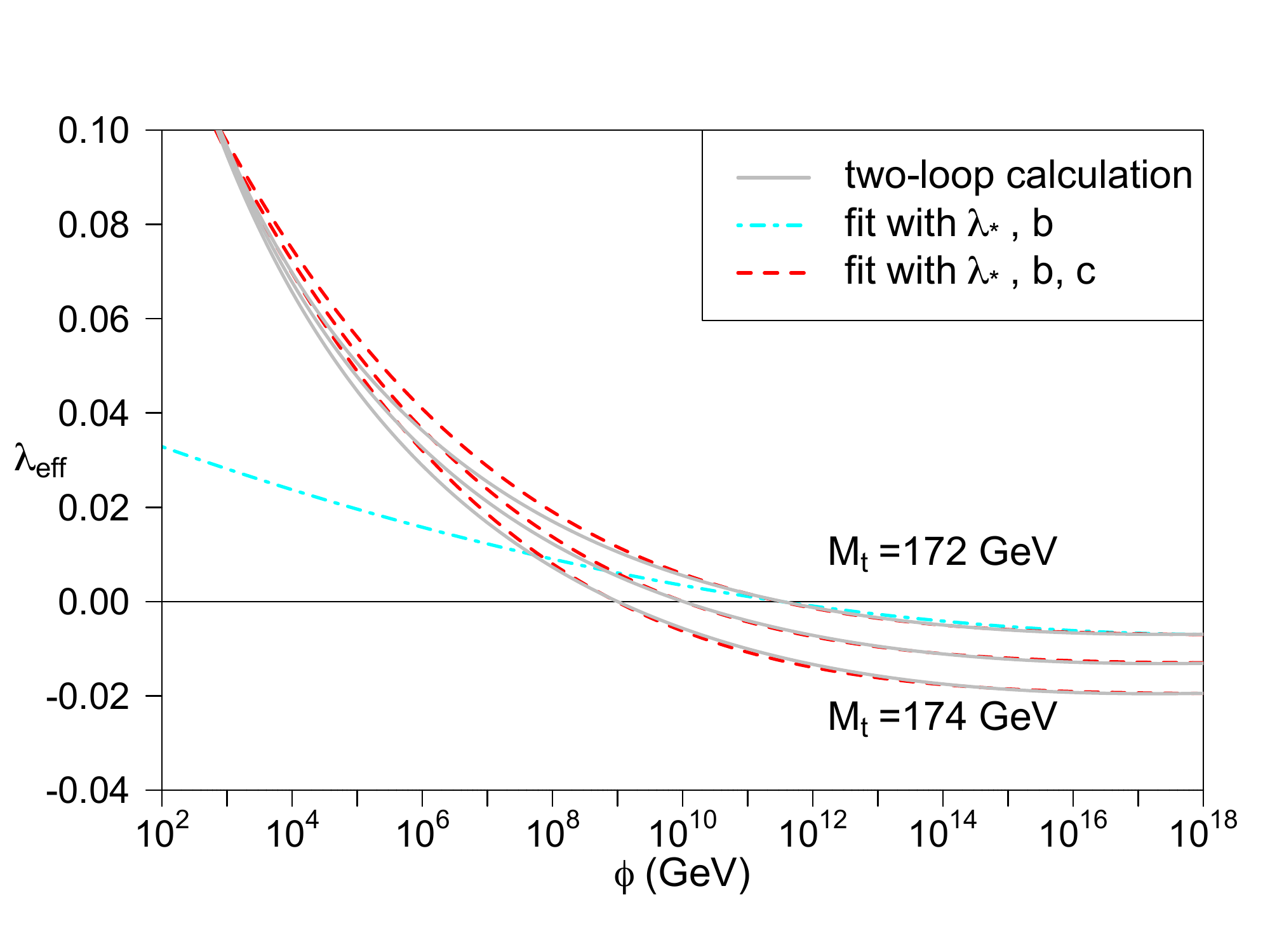}
\caption{
The simplified model of the high-energy effective coupling used for vacuum decay results.
The effective coupling has two free parameters when it is fixed at the lower 
end of the energy range. All three parameters can be fixed by matching to
the Standard Model calculation for a given Higgs and top quark mass.
The plots show Higgs mass $M_{\rm H}=125{\rm GeV}$ and top quark masses
$172{\rm GeV}$ ($\lambda_*=-0.007$),  $173{\rm GeV}$ ($\lambda_*=-0.013$)
and $174{\rm GeV}$ ($\lambda_*=-0.00195$). A two parameter model used 
in earlier work is shown for comparison.}
\label{fig:lambda}
\end{center}
\end{figure}

The precise high energy effective potential 
for the Higgs field has been determined by a two-loop calculation in the context 
of the standard model 
\cite{Ford:1992mv,Chetyrkin:2012rz,Bezrukov:2012sa,Degrassi:2012ry}. 
It is conventionally written in terms of an effective coupling, as
\begin{equation}
V(\phi)=\frac14\lambda_{\rm eff}(\phi)\phi^4.
\end{equation}
The main uncertainty in the potential is due to the uncertainty of the 
top quark mass. The potential has a fairly smooth shape which can be 
computed by direct numerical integration of the $\beta-$functions 
\cite{Espinosa:2007qp}.
Since we are interested in scanning through a range of potentials, 
and exploring the impact of BSM and quantum gravity corrections, 
it is expedient to model the potential analytically by fitting to 
simple functions with a small number of parameters. 
Although two-parameter fits have been used before 
\cite{Degrassi:2012ry,BGM1,BGM2}, we use here a
three parameter model,
\begin{equation}
\lambda_{\rm eff}(\phi)=\lambda_*+b\left(\ln{\phi\over M_p}\right)^2
+c\left(\ln{\phi\over M_p}\right)^4.
\label{higgspot}
\end{equation}
which gives a much better fit over the range of (large) values of
$\phi$ that are relevant for tunnelling phenomena. (See figure
\ref{fig:lambda}.)

Since the value of $\lambda_{\rm eff}$ at energies around the 
Higgs mass is  accessible to experimental particle physics, we 
can fix $\lambda_{\rm eff}$ at the lower end of the range
with some confidence. This leaves two fitting parameters, $\lambda_*$ 
and $b$. We shall explore the dependence of our results on both 
of these parameters, thus our conclusions can be incorporated into 
more general potentials, including non gravitational BSM corrections.

At very high energies, apart from BSM physics, we may have to 
contend with the effects of quantum gravity. We adopt 
the `effective field theory' approach, and add extra polynomial terms to the 
potential which contain the mass scale of new physics,
in this case the Planck mass \cite{Burgess2004,Lalak:2014qua,Loebbert:2015eea}
\begin{equation}
V(\phi)=\frac14\lambda_{\rm eff}(\phi)\phi^4+\frac16\lambda_6{\phi^6\over M_p^2}+\dots
\end{equation}
Adding extra terms to the potential can alter the relationship 
between the original parameters in $\lambda_{\rm eff}$ and the particle masses. 
This is one reason why we will give results in terms of the parameters 
such as $\lambda_*$, rather than top quark or other particle masses.
It is also easier to see how sensitive (or robust) our conclusions are to
the shape of the potential.

\subsection{The ``CDL'' instanton} 

Although Coleman and de Luccia concentrated on the gravitational instanton
representing a bubble with an infinitesimally thin domain wall, the CDL 
instanton is also a good approximation to a wall of finite thickness, as the 
Israel equations are simply a leading order approximation for a thin, but finite 
thickness, wall \cite{Garfinkle:1989mv,Bonjour:1999kz}. As we alter the 
parameters in the potential, the wall can become very thick, to the extent that 
the Higgs may not even reach the true vacuum in the bubble interior. 
The key feature of the CDL instanton is however the O(4) symmetry, 
therefore we refer to an O(4) symmetric configuration of the Einstein-Higgs 
system that has a bubble of lower vacuum energy inside an asymptotically 
flat spacetime as a  ``CDL'' instanton, whether it be a `thin' wall or not.

To find the instanton it is sufficient to consider only a single real component 
of the Higgs field that we denote by $\phi$. The bubble nucleation rate is 
determined by a bounce 
solution with Euclidean metric signature $(++++)$, and action
\begin{equation}
S_E=-\frac{1}{16\pi G}\int_{\mathcal{M}}{\cal R}
+\int_{\mathcal{M}}\left(\frac12g^{ab}\partial_a\,\phi\partial_b\phi+V\right)
\label{4Daction}
\end{equation}
The spacetime geometry should be asymptotically flat with the Higgs field 
at the false vacuum value, and we take the metric ansatz
\begin{equation}
ds^2=d\rho^2+a(\rho)^2\left[d\chi^2
+\sin^2\chi\left(d\theta^2+\sin^2\theta d\varphi^2\right)
\right].
\label{ma}
\end{equation}
The bounce solution $a_b(\rho)$ and $\phi_b(\rho)$ is obtained by solving 
the Einstein-scalar equations,
\begin{align}
&\phi''+{3a'\over a}\phi'-{dV\over d\phi}=0,&\label{phieq}\\
&(a')^2=1+{8\pi G a^2\over 3}\left(\frac12(\phi')^2-V\right).\label{aeq}
\end{align}
The tunnelling exponent is given by the difference in action between the 
bounce solution and the false vacuum. In this case the false vacuum has 
zero action, and the tunnelling exponent is simply $B=S_E[a_b,\phi_b]$.

The tunnelling process is a very high energy phenomenon governed by 
the effective Higgs potential \eqref{higgspot} with the false vacuum at $\phi=0$. 
Requiring solutions which are regular at the origin $\rho=0$ places additional 
conditions on the fields,
\be
\begin{aligned}
&\phi'(0)=0\,,&  &a'(0)=1\,,&\hbox{ at }&\rho=0,\\
&\phi\to0\,,& &a(\rho) \sim \rho \,, &\hbox{ as }&\rho\to\infty.
\end{aligned}
\ee
(In Ref.\ \cite{BGM2} we demonstrated that the condition of metric regularity
could be loosened to allow conical singularities, but the resulting tunnelling rate was 
unaffected.)

Solutions were obtained using a shooting procedure, choosing values of 
$\phi$ at the origin and integrating outwards to find a solution satisfying
the boundary conditions as $\rho\to\infty$. In practice, the boundary conditions 
are applied at some chosen radius $\rho_{\rm max}$, and care has to be taken 
to ensure that the solutions are robust to changes in $\rho_{\rm max}$ 
and $\phi(\rho_{\rm max})$.
\begin{figure}[htb]
\begin{center}
\includegraphics[width=0.6\textwidth]{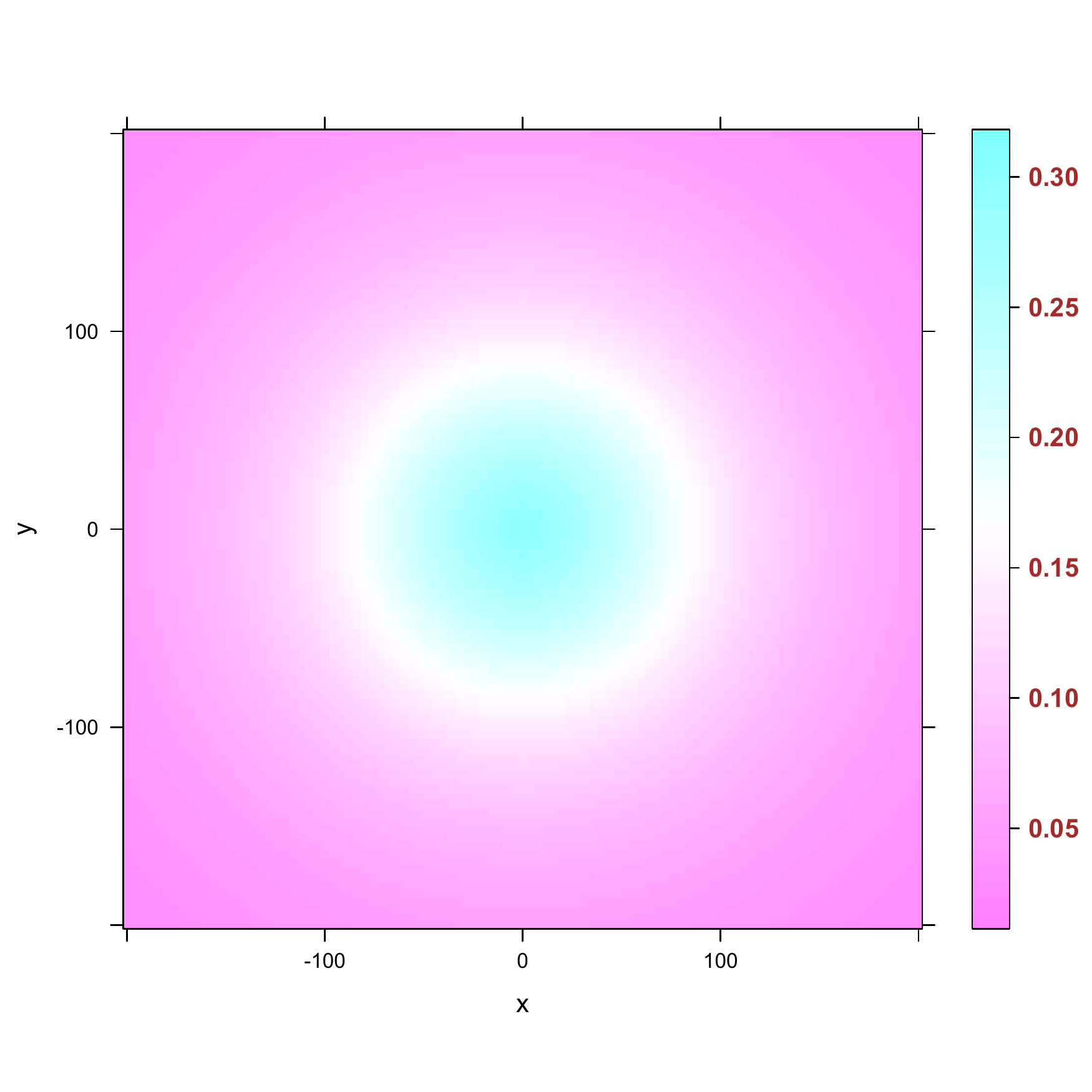}
\caption{The spatial distribution of  the Higgs field in a standard O(4) vacuum 
decay bubble. Two dimensions are shown, the bounce solution has the 
same profile in all four dimensions (three space and one imaginary
time). The central region of the bubble has large values of $\phi$ 
stretching well beyond the potential barrier from the false vacuum 
(pink) into a new Higgs phase (blue).  All measurements are in reduced 
Planck units. The effective coupling here is modelled by 
$\lambda_*=-0.01$, $b=1.4\times10^{-5}$, $c=6.3\times10^{-8}$,
$\lambda_6=0$, corresponding to top quark mass $M_t=173{\rm GeV}$.
}
\label{fig:bounce}
\end{center}
\end{figure}

\begin{figure}[htb]
\begin{center}
\includegraphics[width=0.45\textwidth]{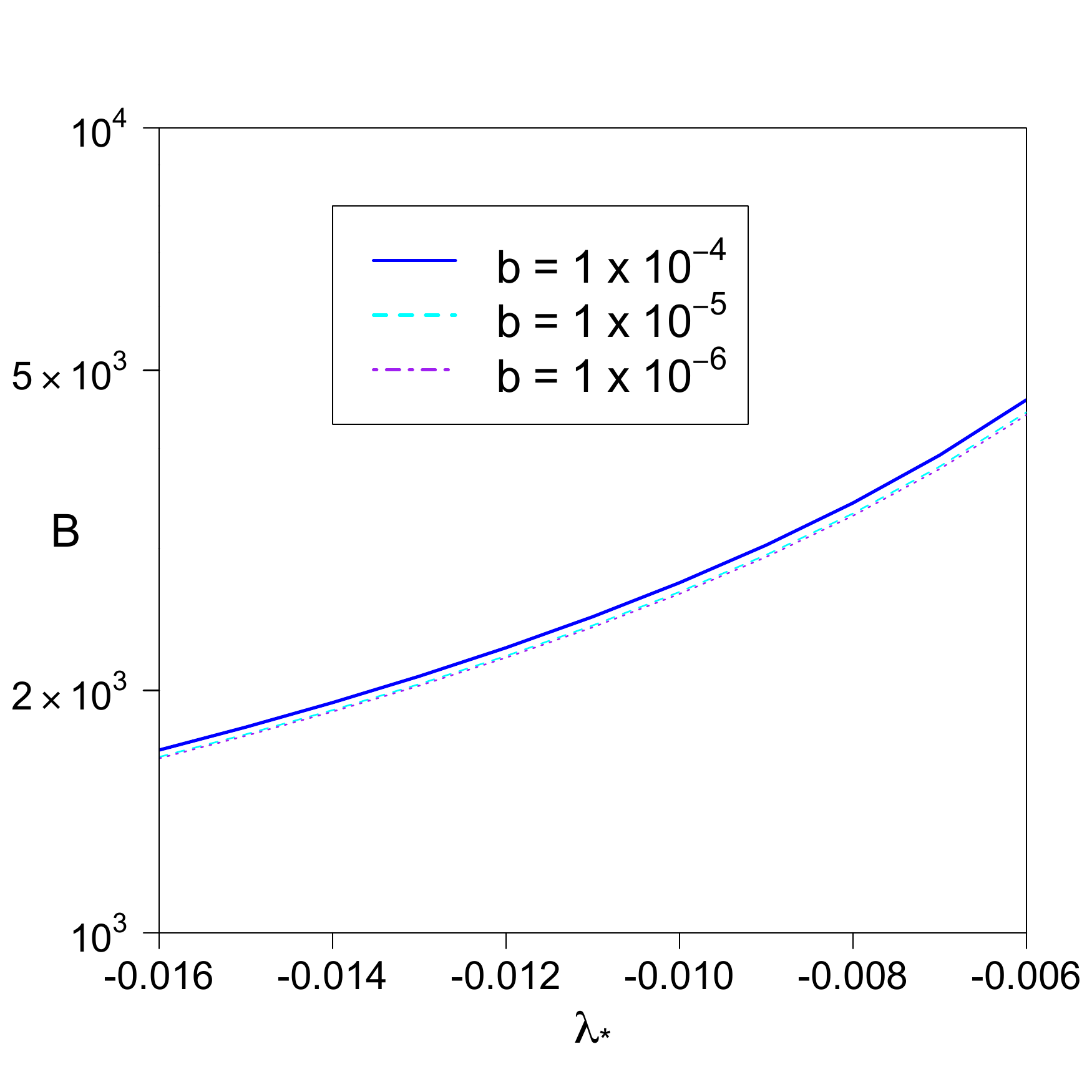}
\includegraphics[width=0.45\textwidth]{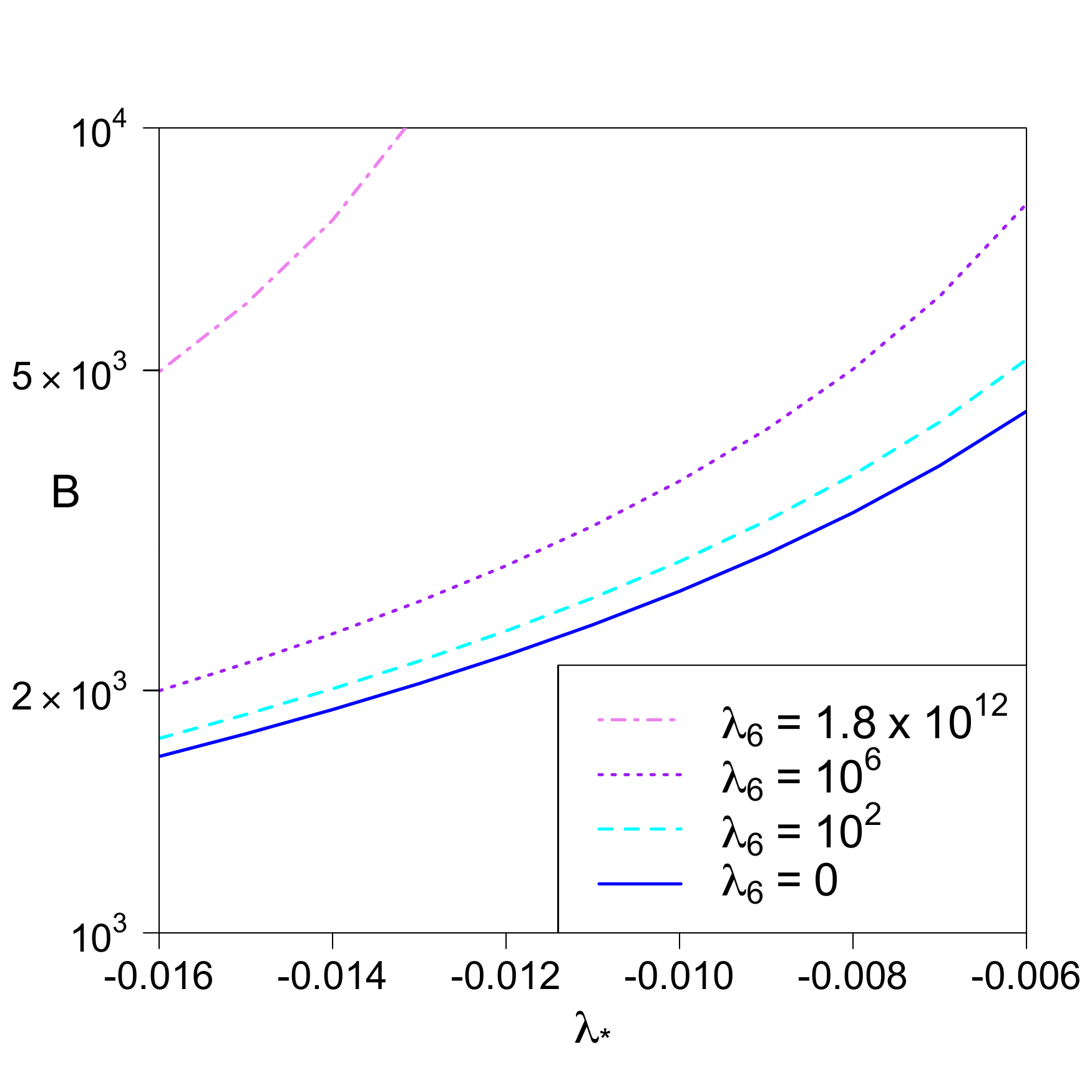}
\caption{The O(4) bounce action $B$ is shown for a variety of Higgs potentials. 
The left panel shows results for $\lambda_6=0$. The principal dependence is 
then on the parameter $\lambda_*$, which determines the large field limit 
of the coupling. There is a very weak dependence on the parameter $b$ as shown. 
The right panel shows the action $B$ as a function of $\lambda_*$ with $b=1.4\times10^{-5}$,
and different values of $\lambda_6$.
}
\label{actionO4}
\end{center}
\end{figure}

An example of the Higgs field for a solution to the Einstein-scalar equation 
without any QG or BSM corrections is shown in Figure \ref{fig:bounce}. 
The centre of the bounce solution 
has negative vacuum energy, and the spacetime geometry around $\rho=0$ 
has negative curvature. The action of the bounce solution is plotted for a range 
of Higgs potentials in figure \ref{actionO4}. The most important dependence 
is on the parameter $\lambda_*$, which varies with the value of the top 
quark mass. There is very little dependence on the $b$ parameter. 

Recall that the tunnelling rate per unit volume is given by $\Gamma_D=Ae^{-B}$. In the 
case where the action includes quantum corrections, the pre-factor is 
determined by the four zero modes which correspond to translations of 
the O(4) symmetric bounce solution. The zero modes contribute $(B/2\pi)^2$ to 
the pre-factor $A$, and there is also a correction from removing the 
zero modes from the effective action. This part is more difficult to calculate, 
but dimensional analysis gives a rough estimate $r_b^{-4}$, 
where  $r_b$ is a characteristic  length scale of the bounce solution. 
For example, the bounce solution in figure \ref{fig:bounce} has $r_b\sim 100M_p^{-1}$.  
To estimate the probability $P_D$ of vacuum decay in the lifetime of the 
universe, we multiply by the volume and age of the observable universe.
We take the size of the universe to be around $10^{61}M_p^{-1}$, leading to 
$P_D\sim \exp(540+2\ln B-B)$, which is comfortably small for the range of 
$B$ values shown in figure \ref{actionO4}.

Now we turn to the effect of physics beyond the standard model, as 
represented by the $\phi^6$ term in the Higgs potential.  Positive values 
of the coefficient $\lambda_6$ increase the height of the potential barrier 
and therefore we expect that this should decrease the vacuum
decay rate. On the other hand, as noted in Refs. 
\cite{Branchina:2013jra,Branchina:2014rva}, 
negative values of $\lambda_6$ should destabilise the false vacuum.

The bounce action for the O(4) symmetric bounce solution with a range 
of values for $\lambda_6$ is shown in figure \ref{actionO4}.
As expected, positive values of $\lambda_6$ increase the action 
and reduce the vacuum decay rate. Negative values of $\lambda_6$
raise the value of $\phi$ at the centre of the bubble to be above the Planck scale
$M_p$. The justification for using the effective field theory
fails, and we cannot confirm enhancement of the tunnelling rate
with the potential and top quark mass ranges we are considering here.

\subsection{Bubble evolution in real time}

The maximal slice of the bounce solution at $\chi=\pi/2$ represents 
a bubble which nucleates at an instant of real time. In the thin-wall case, 
the bubble interior is in the true vacuum, but this is not true for the thick-wall
case. In this section we follow the evolution of the interior towards a final 
state, and see what effect this has on the spacetime geometry.

Following Coleman and De Luccia \cite{CDL}, we perform an
analytic continuation of the bounce solution to Lorentzian spacetime.
The analytic continuation has to be done carefully because, first of all, the metric
is given by a numerical solution and secondly because of the coordinate singularity 
at $\rho=0$. To derive the full bubble interior, we start by choosing a more convenient 
coordinate system $(\tau,r)$ instead of $(\rho,\chi)$,
\begin{align}
\tau&=f(\rho)\cos\chi,\\
r&=f(\rho)\sin\chi.
\end{align}
If we choose $f(\rho)$ to satisfy the equation $f'=f/a$,
with $f(0)=0$ and $f'(0)>0$, then the metric (\ref{ma}) becomes conformally flat,
\begin{equation}
ds^2={a^2\over f^2}\left(d\tau^2+dr^2+r^2(d\theta^2+\sin^2\theta d\varphi^2)\right).
\label{cfe}
\end{equation}
This metric has a very simple analytic continuation to a Lorentzian metric 
with time coordinate $t=-i\tau$,
\begin{equation}
ds^2={a^2\over f^2}\left(-dt^2+dr^2+r^2(d\theta^2+\sin^2\theta d\varphi^2)\right).
\label{cfe}
\end{equation}
The slice of the bounce solution representing the bubble nucleation
which was at $\chi=\pi/2$ is now at $t=0$. The same analytic 
continuation of the metric can be applied to the original $(\rho,\chi)$ 
coordinates by taking
\begin{align}
t&=f(\rho)\sinh\psi_+,\label{tpsi}\\
r&=f(\rho)\cosh\psi_+,\label{rpsi}
\end{align}
where $\psi_+=-i(\pi/2-\chi)$. These relations show that the coordinate transformation 
is only valid for the region $r>t$, covering the exterior of the light-cone centred on the 
point at the middle of the bubble.
Since $\rho$ is unaffected by the analytic continuation, the Euclidean bounce solution 
$\phi_b(\rho)$ becomes an expanding bubble solution 
$\phi(r,t)=\phi_b(\rho)$. Eqs. (\ref{tpsi}) and (\ref{rpsi}) imply
\begin{equation}
\rho=f^{-1}\left[(r^2-t^2)^{1/2}\right].
\end{equation}
Note that, provided $a(\rho)>0$, then $f(\rho)$ is a monotonic function on the 
positive real numbers and the inverse $f^{-1}$ exists. The symmetry under 
Lorentz boosts in $r$ and $t$ is evident. This is the boost part of the full
$O(3,1)$ symmetry.
\begin{figure}[htb]
\begin{center}
\includegraphics[width=0.45\textwidth]{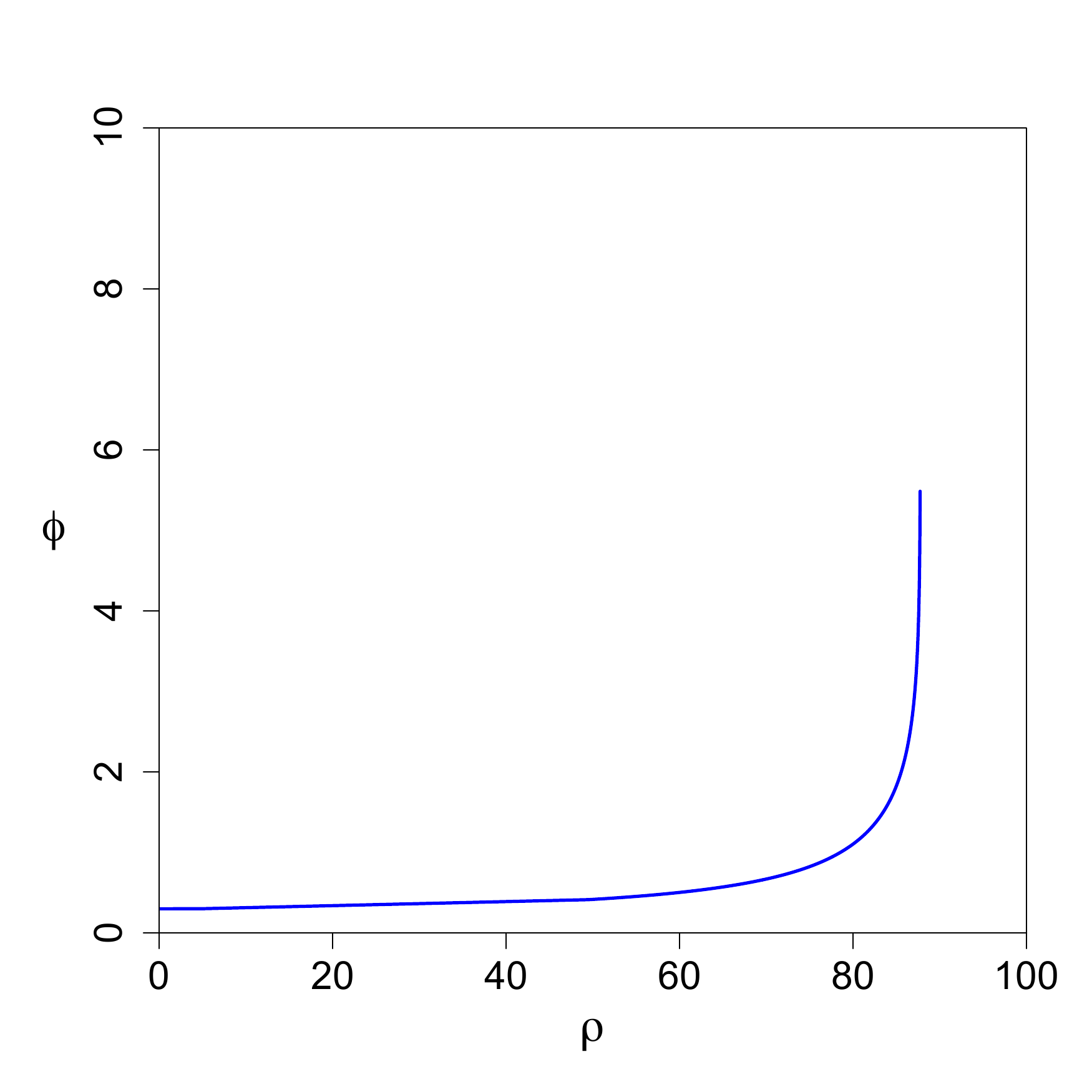}
\includegraphics[width=0.45\textwidth]{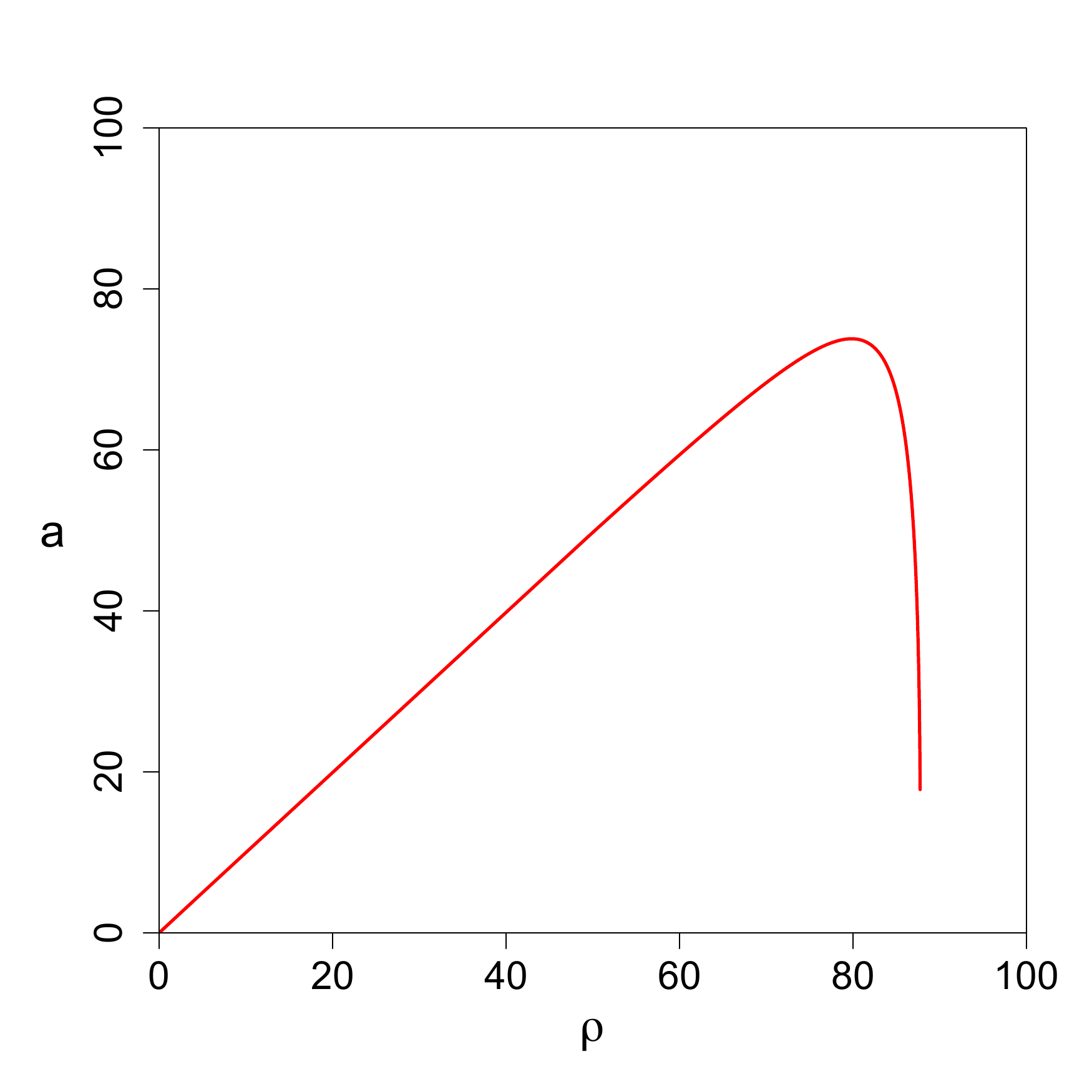}
\caption{
The real-time evolution of the field $\phi$ and the scale factor $a$ inside 
the bubble solution shown in figure \ref{fig:bounce} (which has $\lambda_6=0$).
}
\label{fig:bubble}
\end{center}
\end{figure}

\begin{figure}[htb]
\begin{center}
\includegraphics[width=0.8\textwidth]{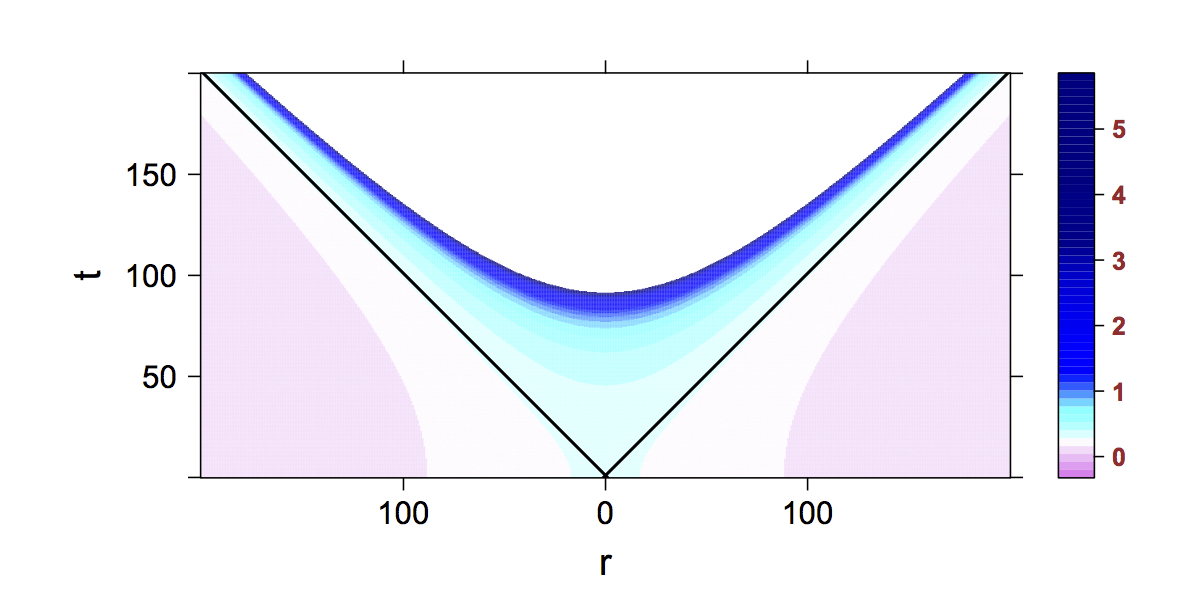}
\caption{
The real-time evolution of the bubble shown in figure \ref{fig:bounce} 
using the conformally flat coordinate system. The lightcone centred on the
bubble is indicated in black.
}
\label{fig:realtime}
\end{center}
\end{figure}

The coordinate system extends trivially through the light cone at $r=t$ and
fixes a set of initial conditions at $\rho=0$ for the evolution of the interior solution,
\begin{equation}
\phi(0)=\phi_b(0),\qquad \phi'(0)=0.\label{incond}
\end{equation}
In the interior $r<t$, we can define a new coordinate system $(\rho_-,\psi_-)$ 
which respects the $O(3,1)$ symmetry of the metric,
\begin{align}
t&=f(\rho_-)\cosh\psi_-,\\
r&=f(\rho_-)\sinh\psi_-.
\end{align}
Again $f'=f/a$, and the interior metric becomes
\begin{equation}
ds^2=-d\rho_-^2+a(\rho_-)^2\left(d\psi_-^2+\sinh^2\psi_-\left(
d\theta^2+\sin^2\theta d\varphi^2\right)\right).
\label{ml}
\end{equation}
The Lorentz symmetry preserves spatial hypersurfaces 
$\rho_-=\hbox{const}$,
and the interior metric in the $O(3,1)$ coordinates is a Friedman metric.
The evolution equations are now Lorentzian versions of \eqref{phieq},
\eqref{aeq}, with initial conditions set on the light cone by Eq. (\ref{incond}).
An interior solution is shown in figures \ref{fig:bubble} and \ref{fig:realtime}.
Unsurprisingly, since the potential in this example reaches large negative
values, the $\phi-$field rolls logarithmically to large values and
the `AdS' spacetime develops a crunch singularity. We see this 
in fig \ref{fig:bubble} as a maximum value of $\rho_-=\rho_s$ where 
$a(\rho_s)=0$ and the kinetic energy of the scalar field diverges. 
For $\lambda_6=0$, the leading order behaviour of the solutions 
when $\rho\approx\rho_s$ can be determined analytically,  
$a\propto (\rho_s-\rho)^{\sqrt{6}/9}$ and $\phi'\propto (\rho_s-\rho)^{-1}$.

\section{Vacuum decay seeded by black holes}

The main aim of this paper is to obtain instanton solutions in the presence
of black holes for general Higgs potentials where the conditions for the thin 
wall approximation break down. We therefore have to solve the fully coupled 
Euclidean Einstein-Higgs equations in the presence of a black hole.
 
First, it will be useful to recall the main conclusions drawn from
the thin-wall approximation calculations described in \cite{BGM1,BGM2}.
There, gravitational instantons were constructed with a false vacuum 
Schwarzschild exterior matched across a domain wall to an exact true 
vacuum AdS (or Schwarzschild-AdS) interior. These Euclidean solutions
exist in principle with all possible values of interior and exterior mass terms,
however, for each seed (exterior) mass black hole, there exists a unique 
least action instanton with a unique remnant (interior) black hole mass.
For very small seed masses, the instanton completely destroys the black
hole, the solution is ``time dependent'' and of the form of a perturbed CDL
instanton. For larger black holes (beyond a critical mass $M_C$,
depending on the vacuum energy and the surface tension of the wall)
the decay process is ``static'' and leaves behind a black hole remnant. 

The critical mass, $M_C$ consists of a static instanton that has no remnant
black hole and is the global minimum in terms of instanton action.
The value of $M_C$ is typically less than the Planck mass for parameters relevant 
to the Higgs potential, therefore to remain within the r\'egime of validity of
euclidean quantum gravity, any reasonable seed mass will be in the
static bounce regime. Our strategy therefore is to numerically construct
static bounce solutions in the expectation that they will dominate the vacuum 
decay rate. Even if these solutions do not have the lowest action, this would only
mean the static instantons constructed would give an upper bound on the
seeded nucleation rate, and our main point about enhancement of the 
decay rate is made a fortiori.

\subsection{Instanton Solutions}

To construct the instanton, we require a geometry with SO(3) invariance and
a Schwarzschild-like mass term; our geometry and scalar field therefore 
depends on a single radial coordinate $r$. It proves numerically 
convenient to take the area gauge, and to parametrise the static, 
spherically symmetric Euclidean metric as:
\begin{equation}
ds^2=f(r)e^{2\delta(r)}d\tau^2+{dr^2\over f(r)}+r^2(d\theta^2+\sin^2\theta d\varphi^2),
\end{equation}
where we write $f$ in the form
\begin{equation}
f=1-{2G\mu(r)\over r}.
\end{equation}
The equations of motion for the bounce solution are therefore
\begin{align}
&f\phi''+f'\phi'+{2\over r}f\phi'+\delta'f\phi'-V_\phi=0,\label{e1}\\
&\mu'=4\pi r^2\left(\frac12f\phi^{\prime 2}+V\right),\label{e2}\\
&\delta'=4\pi G r\phi^{\prime 2}.\label{e3}
\end{align}
Note that by using \eqref{e3} in \eqref{e1}, we can decouple the
equations for $\mu$ and $\phi$, solve, then infer $\delta$ by
integration of \eqref{e3}.

The black hole horizon is defined as usual by the condition $f(r_h)=0$. 
It will be convenient to discuss the solutions in terms of a remnant mass
parameter $\mu_-=\mu(r_h)$, 
rather the actual remnant black hole mass, as in the vicinity of the horizon
we will typically not be in the true AdS vacuum (our Higgs may not have 
fallen to its minimum) nor will our horizon radius be expressible as a
simple ratio of $M_-$. Instead, $r_h=2G\mu_-$ is now a simple ratio
of $\mu_-$, and the expressions in our calculations are much clearer.
The seed mass $M_+$ on the other hand is straightforwardly defined
as the mass at spatial infinity $r\to \infty$, where the field is in the false 
vacuum. Finally, since we integrate out from the event horizon,
it proves convenient to fix the time co-ordinate gauge there,
rather than at asymptotic infinity. This means the $t-$coordinate is
no longer the time for an asymptotic observer, however, the action we
compute is gauge invariant, hence this is irrelevant.

The boundary conditions are therefore
\begin{align}
&\mu(r_h)=\mu_-,\ \delta(r_h)=0,\hbox{ at }r=r_h,\\
&\mu(r)\to M_+,\ \phi(r)\to 0,\hbox{ as }r\to\infty.
\end{align}
If we expand Eqs. (\ref{e1}-\ref{e3}) about the horizon, we obtain a relation 
between $\phi'(r_h)$ and $\phi(r_h)$ which fixes an additional boundary condition,
\begin{equation}
\phi'(r_h)={r_h V_\phi[\phi(r_h)]\over 1-8\pi G r_h^2 V[\phi(r_h)]}.
\end{equation}
This is analogous to the condition $\phi'(0)=0$ in the $O(4)$ case.
The boundary value problem appears to be overdetermined, but this is
simply because the remnant mass parameter $\mu_-$ is determined 
by the value of the seed mass $M_+$. In practise, we solve the system of
equations using a shooting method, integrating from the horizon for a 
given $\mu_-$ and trying different initial values of $\phi(r_h)$. The 
integration leads to an asymptotic value for the seed mass $M_+$ for 
a given remnant mass parameter. From this we can infer the remnant
mass for a given seed mass. 

Before presenting some sample solutions, it is useful to first discuss what
we expect for our functions, using the thin-wall static instantons as a model
solution. Note that the variable $\mu(r)$ includes 
reference to the negative cosmological constant on the true vacuum side:
\be
\mu_{\rm{thin}}(r) = 
\begin{cases}
M_- - r^3/2G\ell^2 &r< r_+\\
M_+ & r \geq r_+
\end{cases}
\ee
where $\ell$ is the AdS curvature radius.
Meanwhile, $\phi(r)$ makes a sharp transition from false to true vacuum at
the static instanton bubble radius, $r_+$. As we move away from the thin wall 
limit, we might expect $\phi$ to be close to its true vacuum value to some
distance outside the horizon before making a more (or less) sharp transition 
to the false vacuum at large $r$, the exception to this behaviour being when 
$\lambda_6=0$, in which case there is no new minimum, and the field will 
simply roll immediately from its maximal value at the horizon to the minimum 
at large $r$. Since $\mu(r)$ responds to the energy-momentum tensor, we 
would expect that as the wall thickens, the sharp jump in $\mu(r)$ at $r_+$ 
will be rounded off and spread out, with the function following the same 
broad shape, but smoothly. As the wall becomes thicker still, the effect of 
the cosmological constant (which makes $\mu$ negative) will become more 
muted, until for the \"uber-thick wall ($\lambda_6=0$) the behaviour of $\mu$ 
will be dominated by the $\phi-$energy-momentum,
and will be mostly positive.

\begin{figure}[htb]
\begin{center}
\includegraphics[width=0.5\textwidth]{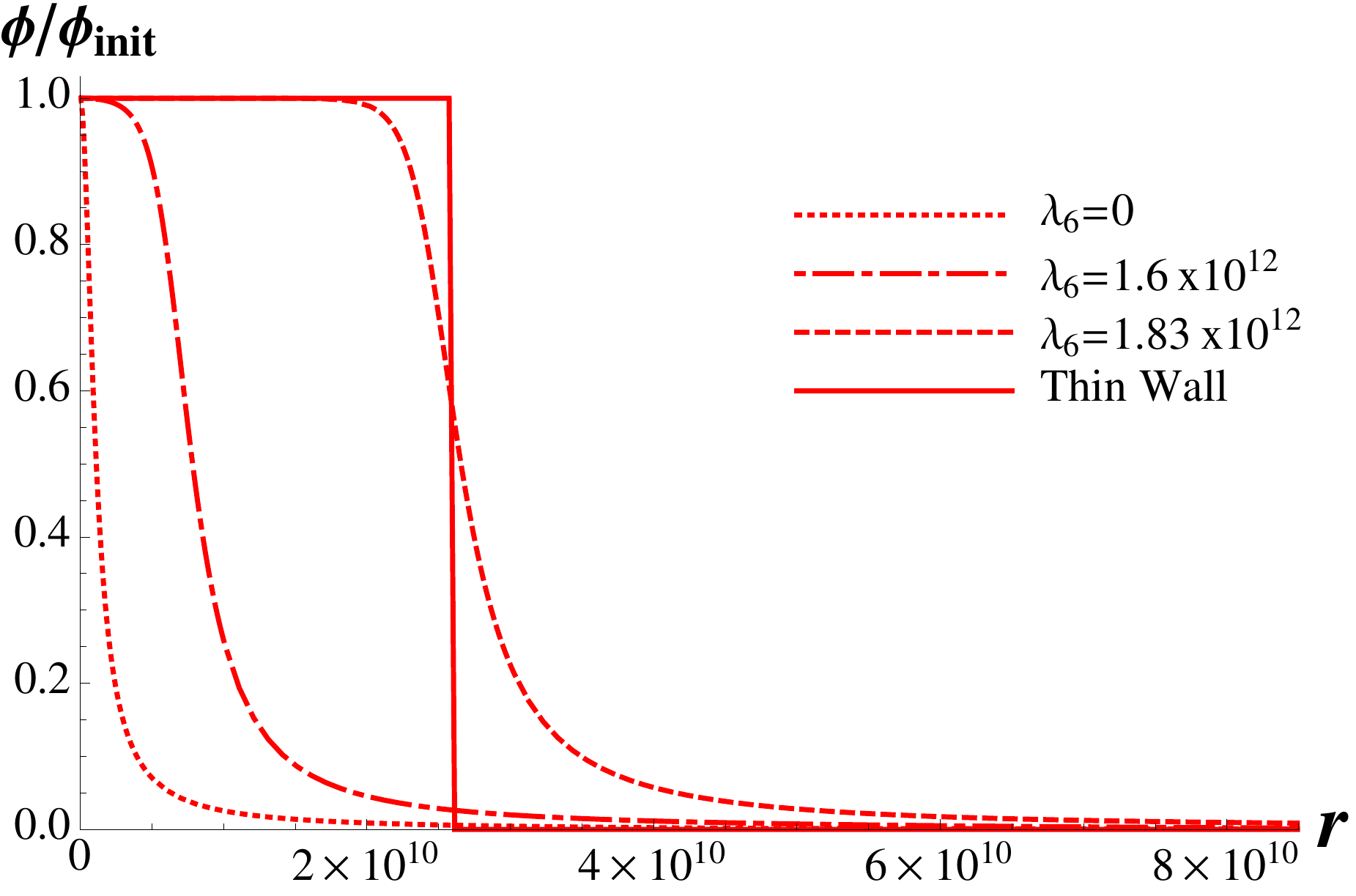}~
\includegraphics[width=0.5\textwidth]{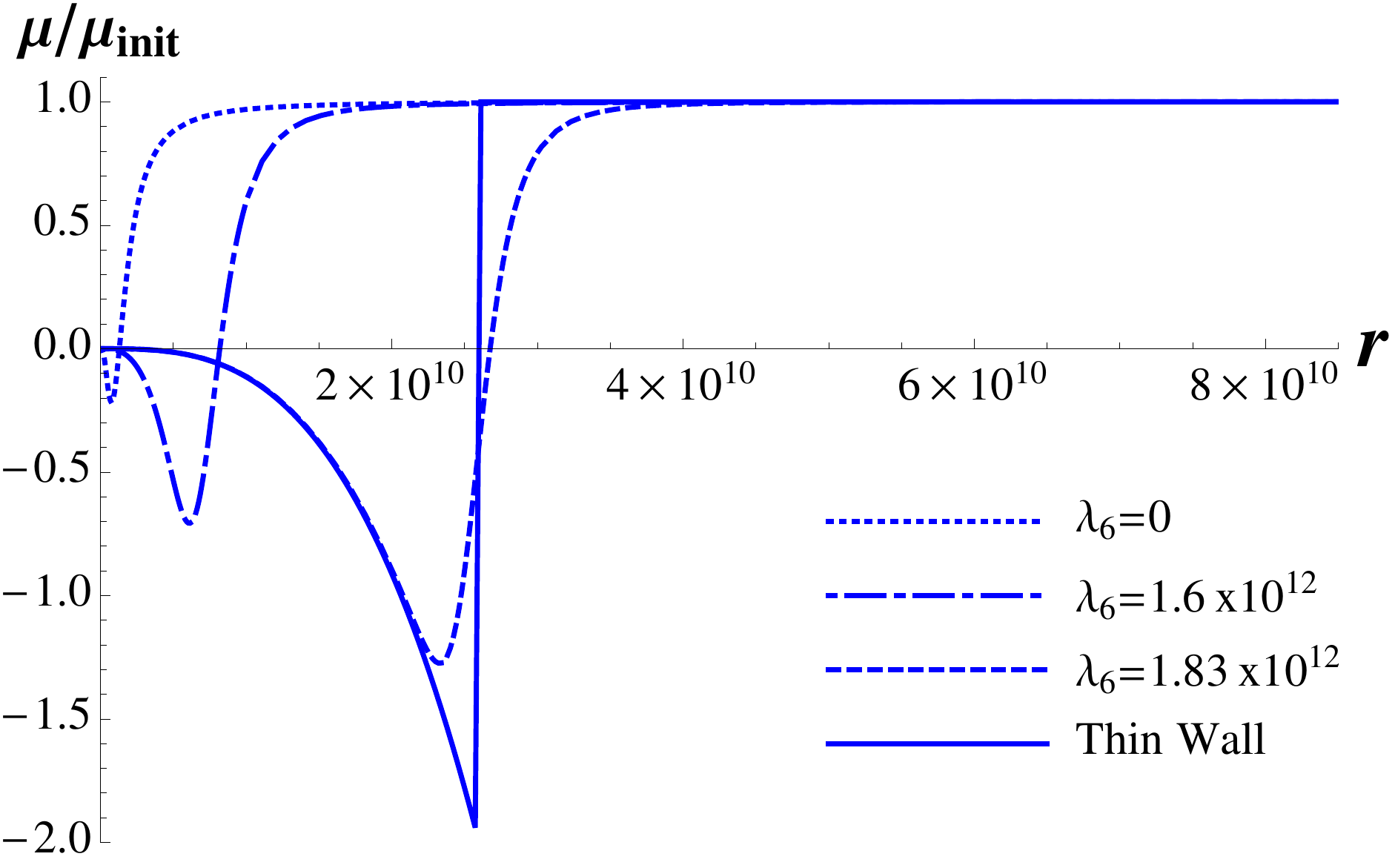}
\caption{
The solutions for $\mu$ and $\phi$ outside the event horizon. Relative
profiles of $\phi$ and $\mu$ are shown, where $\phi$ is shown relative to
its value at the horizon (the maximum) and the $\mu$ function relative
to its asymptotic value, $M_+$.}
\label{fig:profiles}
\end{center}
\end{figure}

Figure \ref{fig:profiles} shows the profiles of the $\phi$ and $\mu$ functions
as the $\lambda_6$ parameter is switched on. For $\lambda_6=0$, there is 
no second minimum of the potential which simply rolls to larger negative
values. We expect therefore that the scalar field will start to roll away from its
horizon value immediately, and the black hole to have a scalar `cloak' where 
the field is rapidly falling to the false vacuum. The $\mu$ profile correspondingly 
is mostly positive, with just a small dip near the horizon where the larger
negative potential has an impact. As $\lambda_6$ is switched on, the `domain
wall' nature of the $\phi-$profile begins to show. In figure \ref{fig:profiles}
an intermediate value of $\lambda_6$ is shown, where the field stays near the
true vacuum in the vicinity of the horizon, but then falls to the true vacuum
over a reasonably thick range of $r$. The geometry function $\mu$ again starts
with the cosmological constant dominated profile, before rising again as the
energy-momentum of the wall causes the mass parameter to change. Finally
the profiles are shown for $\lambda_6$ very close to the thin wall limit. Here, 
we see the $\phi-$profile stays approximately at the true vacuum for a large range
of $r$ near the horizon, then falls relatively rapidly to the false vacuum at large
$r$. The $\mu-$profile tracks the exact Schwarzschild-AdS form until the 
scalar starts to fall, when it makes a rapid transition up to the asymptotic
Schwarzschild form.

\subsection{Computing the action and decay rates}

The $O(3)\times U(1)$ symmetry results in a simple formula
for the tunnelling exponent $B$, derived in Ref.\ \cite{BGM2}:
\begin{equation}
B={{\cal A}_+\over 4G}-{{\cal A}_-\over 4G},
\end{equation}
where ${\cal A}_+$ is the horizon area of the seed black hole and
${\cal A}_-$ is the horizon area of the remnant back hole. 
The action can also 
be expressed in terms of the black hole mass parameters,
\begin{equation}
B={M_+^2-\mu_-^2\over 2M_p^2}.
\end{equation}
We now see why choosing the parameter $\mu$ in the numerical
integration is so convenient -- the tunneling amplitude is simply expressed
in terms of the initial and final values of $\mu$. For a given scalar field 
potential $V$, we can obtain a range of data for different seed masses by
integrating out from the horizon. 

\begin{figure}[htb]
\begin{center}
\includegraphics[width=0.6\textwidth]{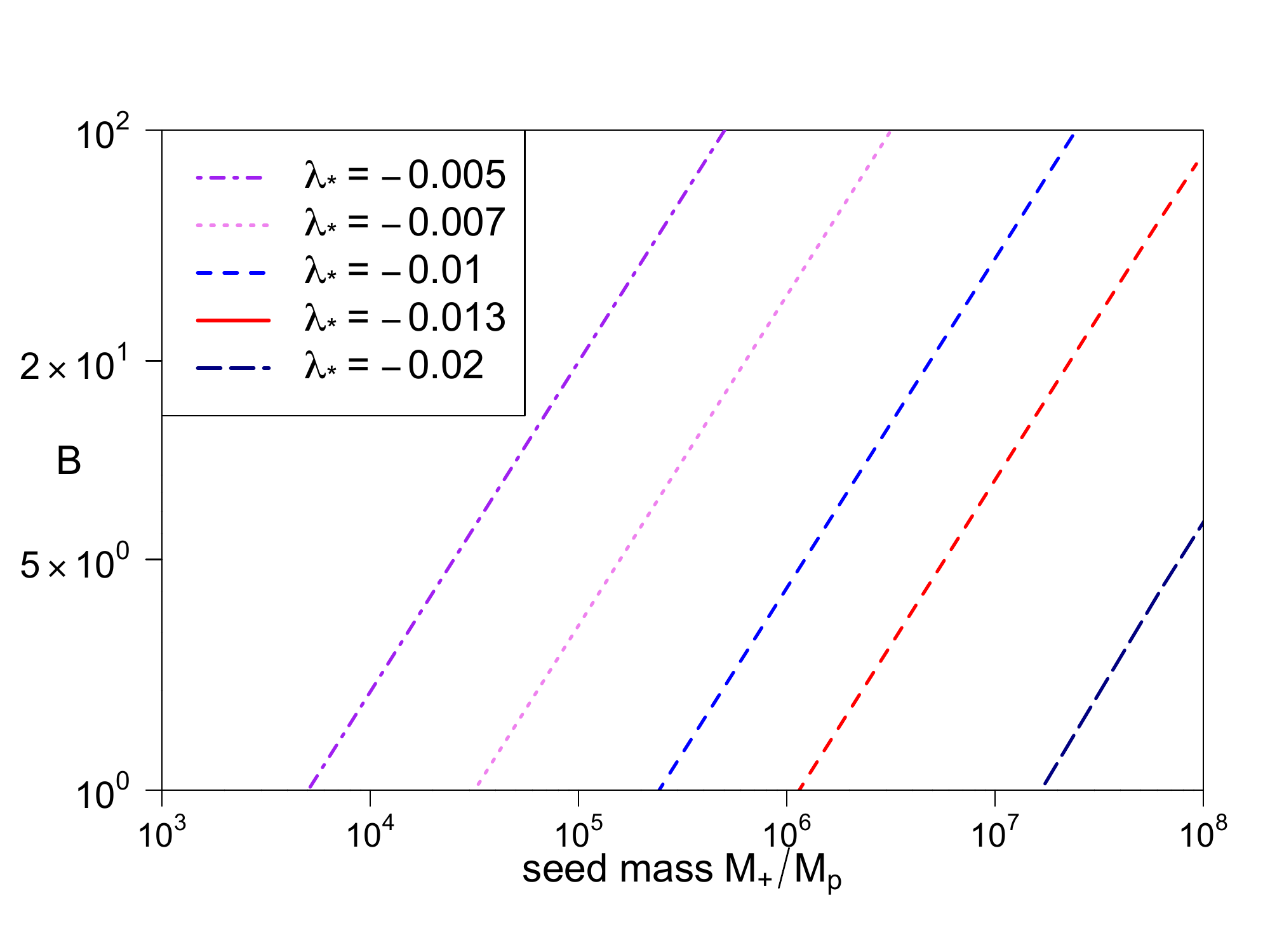}
\caption{
The action of the bounce solution is shown as a function 
of the seed mass for various values of $\lambda_*$.}
\label{fig:actionM}
\end{center}
\end{figure}

Results for the bounce action are shown for different values of the 
seed mass $M_+$ in figure \ref{fig:actionM}. The vacuum decay formalism 
includes a condition that the action $B>1$, and so the plots have been 
restricted to this  range. The plot shows a range of values for the seed mass 
where the bounce action is far smaller than the action of the O(4) solutions
shown in figure \ref{actionO4}. Vacuum decay is enhanced by black holes 
in this mass range.

Given that the seed masses of the black holes favourably
catalysing vacuum decay are rather small, the crucial feature we
have to factor in is whether the vacuum decay is preferential
to Hawking evaporation of the black hole. 
The vacuum decay rate $\Gamma_D$ is given by
\begin{equation}
\Gamma_D=A e^{-B},
\end{equation}
where we have included the pre-factor $A$.  This pre-factor is made up 
from a single factor of $(B/2\pi)^{1/2}$ for the translational zero 
mode of the instanton in the time direction, and a determinant factor. 
\begin{figure}[htb]
\begin{center}
\includegraphics[width=0.48\textwidth]{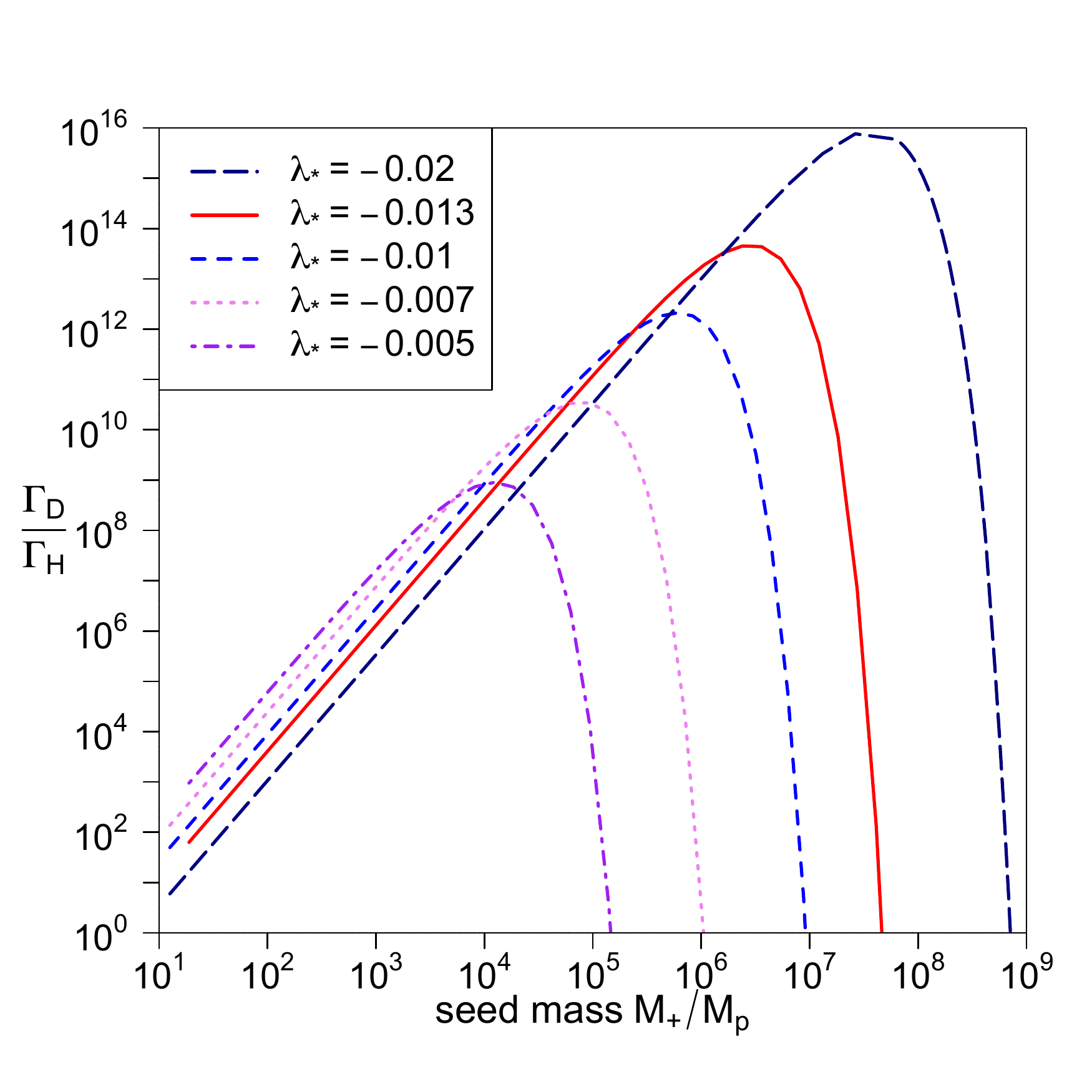}
\includegraphics[width=0.48\textwidth]{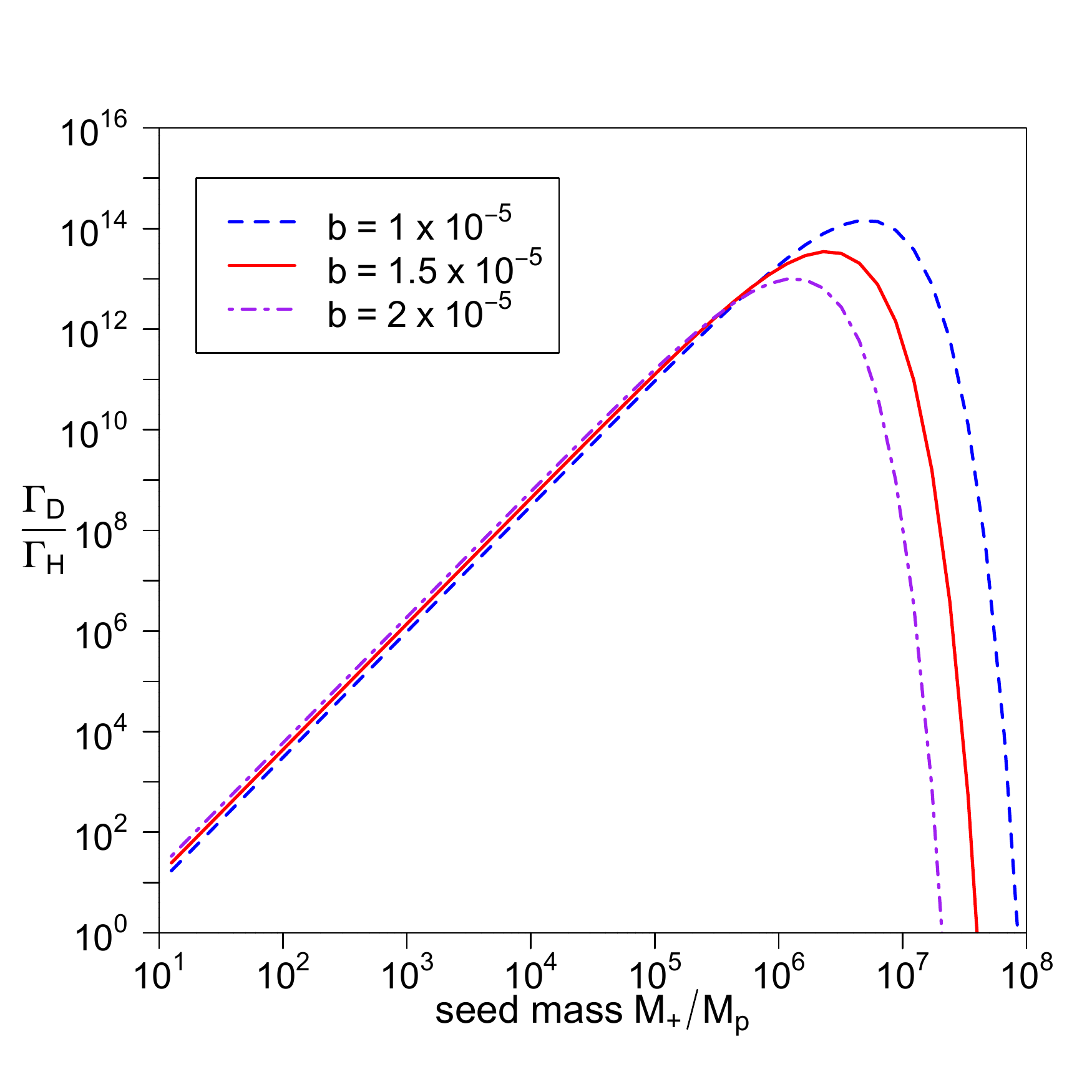}
\caption{
The branching ratio of the false vacuum nucleation rate to the Hawking 
evaporation rate is shown as a function of the seed mass for different 
values of the Higgs potential parameters $\lambda_*$ (with $b=1.4\times10^{-5}$) and $b$
(with $\lambda_*=-0.013$).}
\label{fig:ratio}
\end{center}
\end{figure}
We use dimensional analysis to obtain a rough estimate $(GM_+)^{-1}$
for the determinant factor,  yielding
\begin{equation}
\Gamma_D\approx \left({B\over 2\pi }\right)^{1/2}(GM_+)^{-1}e^{-B}.
\end{equation}
We may use the Hawking evaporation rate 
for a subset of the standard model evaluated by Page 
\cite{Page}. Setting $\Gamma_H=\dot M/M$, we have
\begin{equation}
\Gamma_H\approx 3.6\times 10^{-4}(G^2 M_+^3)^{-1}
\end{equation}
Combining these results, we obtain the branching ratio of the tunnelling 
rate to the evaporation rate as
\begin{equation}
\frac{\Gamma_D}{\Gamma_H}\approx
43.8{M_+^2\over M_p^2}B^{1/2}e^{-B}.
\end{equation}

This branching ratio has been plotted as a function of the seed mass 
$M_+$ for some sets of parameters in figure \ref{fig:ratio}. A primordial 
black hole starting out with a mass around $10^{12}\,{\rm kg}$ would 
decay by Hawking evaporation to the mass scales shown in figure 
\ref{fig:ratio} by the present day.  At some point, the vacuum decay rate 
becomes larger than the Hawking evaporation rate and the black hole 
seeds vacuum decay. The vacuum decay dominates when the black hole 
mass is $10^5-10^9$ times larger than the reduced Planck mass, depending on
where the value of the top quark mass lies in the range $172-174{\rm GeV}$. 
The black holes are large enough for the semi-classical results to be valid, but
with Hawking temperatures in the range $10^{13}-10^{9}\,{\rm GeV}$ their
decay half-life is tiny, ranging from $10^{-24}-10^{-12}\,{\rm s}$.  

\begin{figure}[htb]
\begin{center}
\includegraphics[width=0.6\textwidth]{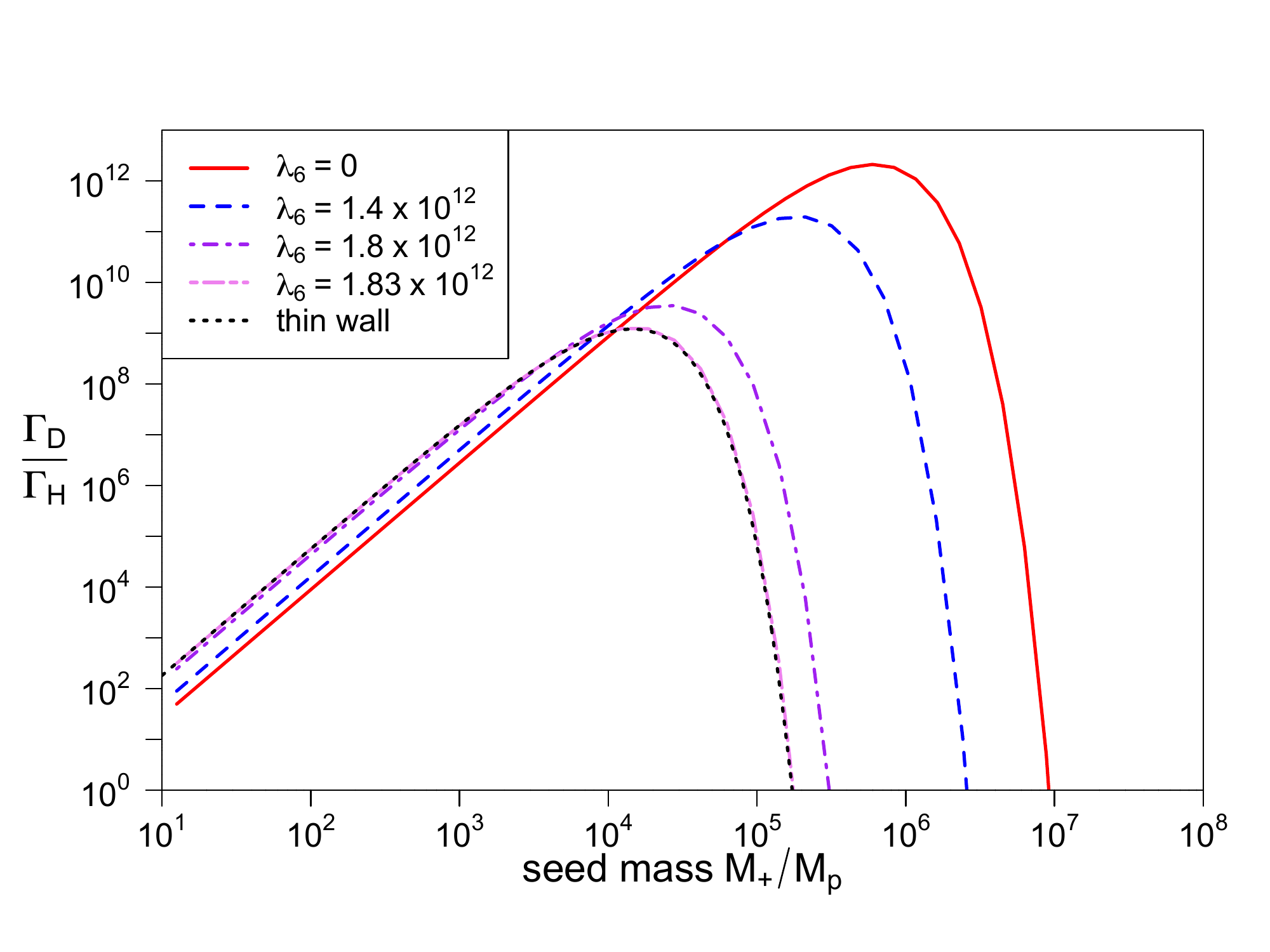}
\caption{
The branching ratio of the vacuum decay rate rate to the Hawking evaporation 
rate as a function of the seed mass with $\lambda_*=-0.01$ and different 
values of the $\lambda_6$ coefficient. Results using a thin-wall approximation are
indistinguishable from the numerical results at the largest value of $\lambda_6$.}
\label{fig:ratioB}
\end{center}
\end{figure}

The effect on the branching ratio of including a $\phi^6$ term in the potential 
is shown in figure \ref{fig:ratioB}. The vacuum decay rate is reduced for positive 
values of $\lambda_6$. As the value of $\lambda_6$ is increased, the 
potential of the true vacuum rises and the bounce solution starts to resemble 
a region of true vacuum surrounded by a thin-wall transition to the false vacuum.  
This allows a cross-check of the numerical results by comparing the bounce 
action to the thin-wall results obtained analytically in Ref.\ \cite{GMW}.

\section{Discussion}

\begin{figure}[htb]
\begin{center}
\includegraphics[width=0.8\textwidth]{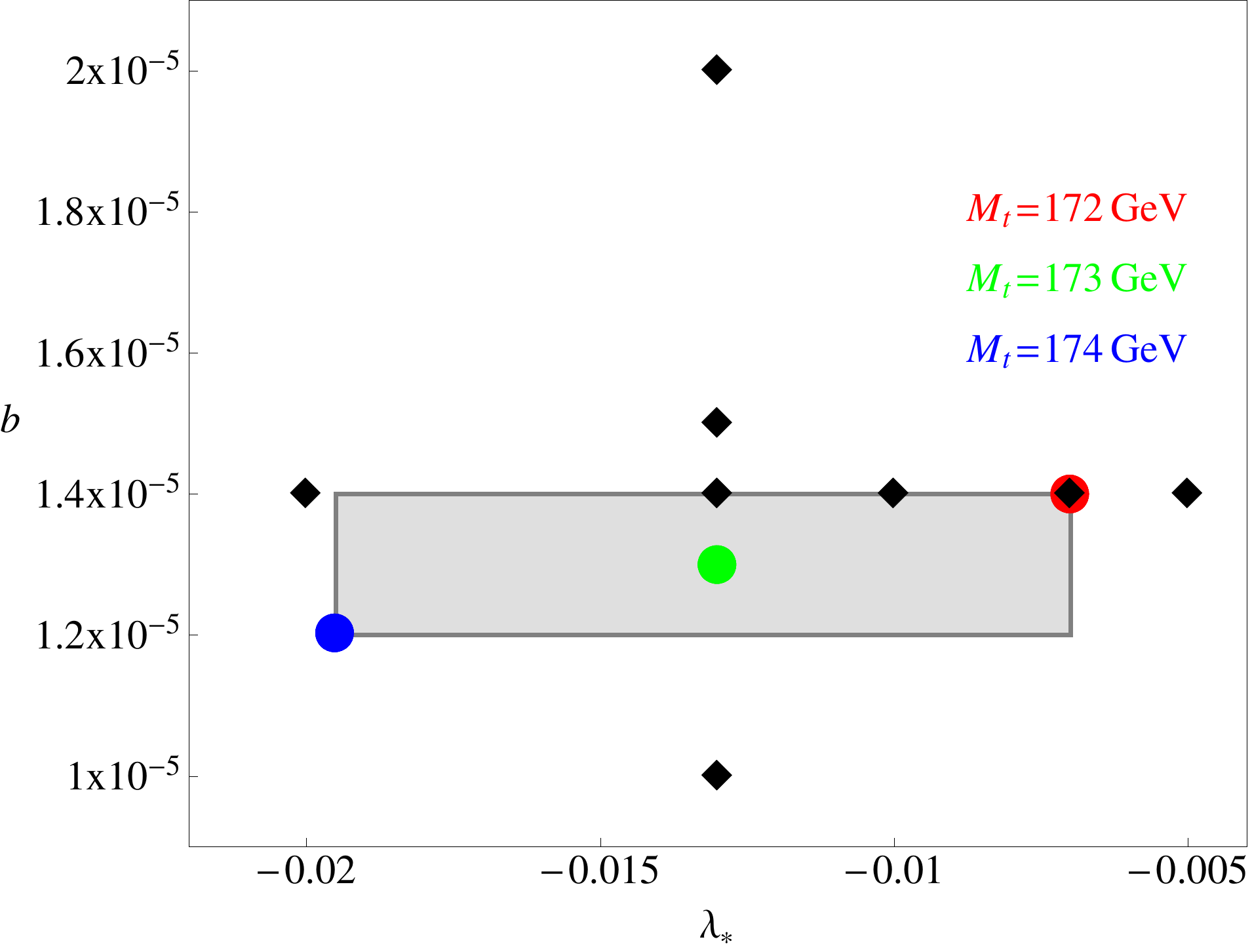}
\caption{
A representation of the parameter space we have explored numerically.
The coloured plot markers represent the parameter values for the
allowed range of top quark mass, $172-174$GeV, and the diamond 
markers the specific parameter values we computed the branching 
ratio for in figure \ref{fig:ratio}. The shaded box represents the parameter
range covered by the Standard Model.
}
\label{fig:paramspace}
\end{center}
\end{figure}

We have shown that our previous result that black holes seed
vacuum decay is extremely robust to the parameters of the Higgs potential.
We used an analytic fit to the Higgs potential and explored a 
range of parameter space beyond that of the Standard Model.
Whereas our previous results applied only to the nucleation of thin-wall bubbles
and covered a very small region of parameter space, these new results apply
for any bubble wall profile and show that black holes
are very effective seeds for vacuum decay. Figure \ref{fig:paramspace}
shows the region of parameter space explored vs.\ the 
standard model parameter range.

The importance of these results lies in the fact that a single primordial black
hole in the observable universe would cause the decay of the Standard
Model Higgs vacuum, and therefore would contradict the Standard Model.
Looking beyond the Standard Model, quantum gravity effects can suppress 
the vacuum decay rate by contributing $\phi^6$
terms to the Higgs potential, but the vacuum decay rate still remains large
unless the high-energy vacuum becomes the false vacuum, which happens
when the coefficient $\lambda_6$ is around $10^{12}$.
A stable Higgs vacuum requires the new physics to change the barrier
in the Higgs potential at energies around $10^{10}-10^{14}{\rm GeV}$.

Vacuum decay can also be enhanced if the $\lambda_6$ coefficient
in the potential is negative. However, we have found that the Higgs field
at the centre of the vacuum decay bubble lies very close to the Planck scale
and the reliability of the effective potential becomes questionable
for negative values of $\lambda_6$. For non-negative values of
$\lambda_6$, vacuum decay rates for unseeded vacuum decay bubbles 
are extremely small. Nevertheless, we have found a way to examine the 
evolution of the bubbles in real time and followed the interior towards a 
singularity.

Bubble nucleation in the  presence of a black hole raises a number of
questions which should be investigated further. The instanton approach, 
and its interpretation, are based on results which well understood in
flat spacetimes but not rigorously described so far in the curved space 
context (although see \cite{Rubakov:1999ir,Brown:2007sd}). 
One question is the role of Hawking
radiation in the tunnelling process. We have shown that the thermal 
evaporation rate is negligible, but there are still questions about the
global spacetime structure, and why the result for the tunnelling rate
is independent of the angle in the conical singularity arising in the
instanton \cite{GMW}. There is also a question about
taking into account the way in which quantum corrections to the potential are
affected by the spacetime curvature, although to some extent this question can
be side-stepped by looking at black hole monopoles where the charge
can be used to reduce the Hawking temperature, as was done in
\cite{BGM2}. There are also a variety of interesting other consequences
of finite temperature tunnelling, particularly in a cosmological context,
see for example \cite{Greenwood:2008qp,Cheung:2013sxa,Espinosa:2015qea}.

Besides primordial black holes, another source of nucleation seeds could
be black holes formed by particle collisions in theories with a low fundamental 
Planck mass 
\cite{ArkaniHamed:1998rs,Antoniadis:1998ig,Randall:1999ee,
Kanti:2004nr,Gregory:2008rf}. 
The possibility of vacuum decay caused by
black holes formed from collisions was considered in \cite{BGM1,BGM2}.
There is an observations constraint here due to long life of our vacuum state
despite the existence of high energy cosmic ray collisions, which may place
interesting limits on theories with a low fundamental Planck mass.

Finally, although we have considered bubbles inside a Schwarzschild 
(i.e.\ asymptotically flat) spacetime, AdS-AdS transitions, (such as 
considered in \cite{Sasaki:2014spa} to address the information problem) 
are obviously of interest. Static bubbles would now have the holographic 
interpretation of flows between
field theories at different temperatures and different central charges.
Flows and bubbles in AdS have of course already been considered,
but the new aspect of having a black hole raises intriguing possibilities
for thermal flows.

\section*{Acknowledgements}
We would like to thank Erick Weinberg for useful discussions and
Joan Elias-Mir\'{o} for helpful correspondence.
PB was supported in part by an EPSRC International Doctoral Scholarship,
and by the Einstein Research Project ``Gravitation and High Energy Physics'', 
funded by the Einstein Foundation Berlin, the Israel Science Foundation 
grant no.\ 812/11 and by the Quantum Universe grant from the I-CORE 
program of the Israeli Planning and Budgeting Committee.
RG and IGM are supported in part by STFC (Consolidated Grant ST/J000426/1).
RG is also supported by the Wolfson Foundation and Royal Society, and
Perimeter Institute for Theoretical Physics. 
Research at Perimeter Institute is supported by the Government of
Canada through Industry Canada and by the Province of Ontario through the
Ministry of Research and Innovation. 
RG would also like to thank the Aspen Center for Physics for hospitality.
Work at Aspen is supported in part 
by National Science Foundation Grant No.\ PHYS-1066293 and the hospitality 
of the Aspen Center for Physics.

\providecommand{\href}[2]{#2}
\begingroup\raggedright\endgroup

\end{document}